\documentclass[11pt]{article}
\usepackage{multicol}
\usepackage{booktabs}
\usepackage{amssymb,bm,mathrsfs,bbm,amscd}
\usepackage[tbtags]{amsmath}
\usepackage{lastpage}

\title{ Scaling Law of Beam Break-Up for the Single Ultrashort Bunch in RF Linac }

\author{Xiongwei Zhu \\
{Institute of High Energy Physics, Chinese} \\
{Academy of Sciences, Beijing 100049, China} }

\begin{document}
\maketitle

\begin{abstract}
   Femtosecond bunch is a hot topic in the present world accelerator research community. The high peak current bunch will
lead to the beam breakup phenomenon. In this paper, the scaling law with current for the single ultrashort bunch beam breakup in the radio
frequency linear accelerator is proposed and obtained. The scaling power factor is roughly estimated to be $1/6$, $1/6$ and $1/3$ for
the geometry wake, the surface roughness wake and the resistance wake respectively.

\end{abstract}



\section{Introduction}

\hskip 12pt   Beam breakup ( BBU )\cite{4} is a very popular subject in the radio frequnecy linear accelerator community, particularly for the high current relativistic electron beam. Collective instability arises for the bunched beam in the radio frequency linac, the pulsed beam in
the induction linac, in the large-scale free-electron laser, and in the international linear collider. Many methods have been
proposed to control BBU. With the developments of the X-ray free electron laser ( XFEL )\cite{1,2,3}and the international linear collider ( ILC ), there is a trend to study the femtosecond bunch technique. The present level of our electron gun can not satisfy the requirements of XFEL and ILC. Usually we use the bunch compressor\cite{1,2} to get the femtosecond bunch. The peak current increases very much in bunch compressing and will lead to the beam breakup of the short bunch. In the past, A.Chao\cite{a1}, B.Richter and C.Y.Yao first set up the basic equation for the transverse beam dynamics in RF Linac. Yokoya\cite{a}, Y.Y.Lau\cite{a2}, J.Delayen\cite{c3,c2}, Zimmermann\cite{b}, K.Y.Ng\cite{c}, J.-M.Wang\cite{c1}, and Whittum\cite{5} have done some analytical works about the single bunch and the multibunch BBU. In this paper, we discuss the single ultrashort bunch BBU and propose to give an approximate scaling law with the beam current in such machine for the single ultrashort bunch BBU.

\section{Tranverse motion equation for the single bunch BBU}

\hskip 12pt Beam break up ( BBU ) phenomenon is an old research topic in RF linac field and was first observed in 1957\cite{4}. BBU is related to the dipole mode in the accelerating structure. BBU is also called the beam pulse shortening phenomenon and often occurs in the multibunch
operation mode. The tail bunch in the bunch train is deflected by the dipole mode excited by the front bunch in the bunch train and is blown out. While, in case of the single bunch BBU, the tail part of the bunch is deflected by the head part of the bunch.

   In case of the single bunch, the transverse motion equation\cite{a1} for the beam centroid $X(z,\xi)$ is expressed as
\begin{equation}
  (\frac{\partial}{\partial z} \gamma
\frac{\partial}{\partial z} + \gamma k^{2})X(z,\xi) =
\int_{0}^{\xi} d \xi' W(\xi - \xi')\nu(\xi') X(z,\xi'),
\end{equation}
where $\xi = t - z/c $ is the distance along the beam and varies from zero at the head to the beam pulse length, $t$ is the
time, $z$ is the longitudinal position along the linac, $c$ is the light velocity, $W(\xi)$ is the point Green
wake function, $\nu = I/I_{A}$ is the Budker parameter, $I$ is the beam current, and $I_{A}$ is the Alfven current.

\section{ The point charge wake( Green function ) for the ultrashort bunch }

\hskip 12pt   The charged particle beam interacts with its vacuum chamber surroundings electromagnetically. The electromagnetic field left behind by the bunch is
called wakefield\cite{6,8}. The wakefield of the short bunch is difficult to get, and is extensively studied. Usually, we first get the wakefield of the point charge, and then
use the integration to get the wakefield of the bunch made of point charges. In this paper, we consider three kinds of wakefields: the geometry wakefield, the surface roughness wakefield and the resistance wakefield. The transverse geometry wakefield at zero point is estimated as
\begin{equation}
W^{'}(0)=\frac{2Z_{0}c}{\pi a^{4}}.
\end{equation}
In case of the ultrashort bunch, we can cut Taylor expansions to the first order to have a well approximate point Green geometry wake
function\cite{6} as
\begin{equation}
W_{1}(\xi)=W^{'}(0) \xi=\frac{2Z_{0}c}{\pi a^{4}} \xi,
\end{equation}
where $Z_0$ is the free space impedance, $c$ is the light velocity and $a$ is the vaccumm chamber radius.

As for the surface roughness wakefield\cite{b}, the point Green longitudinal wake caused by the dieletric layer or the surface roughness is
\begin{equation}
W_{sl} = \frac{Z_{0} c}{\pi a^{2}} cos( k_{0}s ),
\end{equation}
where $k_{0} = (\frac{2 \varepsilon}{a \delta ( \varepsilon - 1)})^{\frac{1}{2}}$, $\delta$ is the width of the layer, $\varepsilon=2$ for the surface roughness. The corresponding longitudinal impedance is then
\begin{equation}
Z_{L} = \frac{Z_{0} c}{\pi 2 a^{2}} (\delta (k - k_{0}) + \delta (k + k_{0})).
\end{equation}
Using Panofsky-Wenzel theorem, we have $Z_{T}(k) = 2 Z_{L} (k) / k a^{2}$. The transverse point Green wake is
\begin{equation}
W_{st} = \frac{Z_{0}c}{\pi a^{4}} \frac{sin (k_{0}s)}{k_{0}}.
\end{equation}
In case of the ultrashort approximation, the point Green surface roughness wake is also well approximated as
\begin{equation}
W_{2}(\xi)=W^{'}(0) \xi=\frac{Z_{0}c}{\pi a^{4}} \xi.
\end{equation}
In comparision with the geometry case, formula (7) is half of the formula (3) and has the same form.

 Finally, we use the following transverse resistant point Green wake\cite{8} in case of the short bunch roughly,
\begin{equation}
W_{3}(\xi) = \frac{Z_{0} c}{4 \pi^{2} a^{3}} \sqrt{\frac{c}{\sigma}} \frac{1}{\sqrt{\xi}},
\end{equation}
where $\sigma$ is the conductivity. Actually, the Green function near the zero point has the transient oscillating characteristic.
We adopt the solution above from the negative direction along the coordinate axis.

\section{Scaling law}

\hskip 12pt  In case of the geometry wakefield,  we use the point Green wake $W_{1}(\xi)$ and suppose
\begin{equation}
 X(z,\xi)=Im{(\chi(z,\xi)exp(ik (\xi) z)}.
\end{equation}
Using the simple linear accelerating case $\gamma =\gamma_{0}+ Gz$ ( G is the accelerating gradient ), we can get the equation for $\chi(z,\xi)$
\begin{equation}
( G+2ik \gamma )\frac{\partial \chi}{\partial z} + ikG \chi = \int_0^{\xi} d \xi' W(\xi-\xi') \nu( \xi' ) \chi.
\end{equation}
We take the Laplace transformation of $\chi$ to $\tilde{\chi}=L[\chi]$ and get the following equation
\begin{equation}
(G+2ik \gamma) \frac{\partial \tilde{\chi}}{\partial z}+ ikG \tilde{\chi}= \nu \tilde{W }(p) \tilde{\chi}(z,p).
\end{equation}
The solution of $\tilde{\chi}$ is
\begin{equation}
\tilde{\chi}= \tilde{\chi} (0,p) exp(- \int\frac{\nu \tilde{W}(p) - ikG}{G+2ik \gamma} dz ),
\end{equation}
where $\tilde{\chi}(0,p)=\int_0^{\xi'}exp(-p \xi') d \xi'$. Then, we get the solution of $X$ as
\begin{equation}
X(z,\xi) = \int_c dp \tilde{\chi} (0,p) exp(\theta (p)),
\end{equation}
where $\theta = p \xi - \frac{\nu ln(f(z))}{ikG} \frac{Z_0 c}{\pi a^4} \frac{1}{p^2} + \frac{1}{2} ln(f(z)), f(z) = 2 i k G z + G + 2 i k \gamma_0 $.
From $\theta' (p) = 0 $, we get $ p = ( - \frac{\nu ln(f(z))}{ik G } \frac{2 Z_0 c}{\pi a^4 \xi})^\frac{1}{3}$, and
\begin{equation}
\theta^{(2)} = 2.38 (- \frac{i k G}{\nu ln(f(z))})^{\frac{1}{3}} (\frac{\pi a^4}{Z_0 c})^{\frac{1}{3}} {\xi}^{\frac{4}{3}}.
\end{equation}
Finally we use the steepest descent method to obtain
\begin{equation}
X(z,\xi)\approx \tilde{\chi} (0,p) (2 \pi \theta^{(2)})^{- \frac{1}{2}} exp \theta.
\end{equation}
After some managements, we can write a simple scaling law formula from the complex expression as
\begin{equation}
X=b (\frac{I}{I_A})^{\frac{1}{6}},
\end{equation}
where $b$ is a constant and can be determined by the machine study. From the formula above, we can see that $X$ increases with the increasing of the bunch current $I$, and the scaling power factor is estimated to be $1/6$ roughly. We note that Yokoya\cite{a} included the same power factor in a different complex expression. BBU occurs when $X$ is bigger than the beam pipe radius.

Further, in case of the surface roughness wakefield, we use the point Green wake $W_{2}(\xi)$ and will get the similiar result with the threshold value half of the geometry wakefield case. The scaling power factor is also estimated to be $1/6$.

Finally,  in case of the resistance wakefield, using the point wake $W_{3}(\xi)$ roughly, we obtain
\begin{equation}
\theta = p \xi - \frac{\nu ln(f(z))}{2ikG} \tilde{W}(p) + \frac{1}{2} ln(f(z)),
\end{equation}
\begin{equation}
\tilde{W}(p) =  \frac{Z_{0} c}{4 \pi^{2} a^{3}} \sqrt{\frac{c \pi}{\sigma}} \sqrt{\frac{1}{p}}.
\end{equation}
From $\theta' (p) = 0 $, we get $ p = ( -\frac{1}{2} \frac{\nu ln(f(z))}{2ik G } \frac{Z_0 c}{4 \pi^{2} a^3 \xi} \sqrt{\frac{c \pi}{\sigma}})^\frac{2}{3}$, and the second derivative $\theta''(p) = - \frac{3}{4} \frac{\nu ln(f(z))}{2 i k G} \frac{Z_{0} c}{4 \pi^{2} a^{3}} \sqrt{\frac{c \pi}{\sigma}} p^{-\frac{5}{2}}$. We can also get from formula (15)
\begin{equation}
X=b_{1} (\frac{I}{I_A})^{\frac{1}{3}},
\end{equation}
$b_{1}$ is a constant related to the machine study. Then, the scaling power factor is roughly estimated to be $1/3$  in case of the resistance wakefield.

\section*{Acknowledgement}
\hskip 12pt Thank F.Zimmermann ( CERN ) and J.Delayen ( JLab ) for the useful discussions.


\begin{thebibliography}{4}
\bibitem{4} {M.C.Kelliher, R.Beadle, Pulse-shortening in Electron Linear Accelerators, Nature, 187, 1099 ( 1960 ).}
\bibitem{1} {Linac Coherent Light Source Design Report, SLAC-R-593 ( 2002 ); TESLA Technical Design Report ( 2001 ); SCSS XFEL CDR
( 2005 ).}
\bibitem{2} {Xiongwei Zhu, et al, Layout of Bunch Compressor for Beijing XFEL Test Facility, NIM A, 566, 250 ( 2006 ).}
\bibitem{3} {Xiongwei Zhu, Ultrashort High Quality Electron Beam from Laser Wakefield Accelerator using Two-step Plasma Density Profile, Review of Scientific Instruments, 81, 033307 ( 2010 ).}
\bibitem{a1}{A.W.Chao, B.Richter, C.Y.Yao, Beam Emittance Growth Caused by Transverse Deflecting Fields in a Linear Accelerator, NIM, 178,1
( 1980 ).}
\bibitem{a} {K.Yokoya, Cumulative Beam Breakup in Large-scale Linacs, Desy 86-084 ( 1986 ).}
\bibitem{a2}{Y.Y.Lau, Classification of Beam Breakup Instabilities in Linear Accelerator, PRL, 63, 1141 ( 1989 ).}
\bibitem{c3}{J.Delayen, Cumulative beam breakup in linear accelerators with arbitrary beam current profile, PRST-AB, 6, 084402 ( 2003 ).}
\bibitem{c2}{J.Delayen, Cumulative Beam Breakup in Linear Accelerators with Time-dependent Parameters, PRST-AB, 8, 024402 ( 2005 ).}
\bibitem{b}{F.Zimmermann, D.Whittum, C.K.Ng, M.E.Hill, Wake Fields in a mm-wave Linac, AIP Conference Proceedings 472, 270, New York
( 1998 ).}
\bibitem{c} {K.Y.Ng, C.L.Bohn, Theory of Cumulative Beam Breakup with BNS Damping, APAC01, 372 ( 2001 ).}
\bibitem{c1}{J.-M.Wang, J.Wu, Cumulative Beam Breakup due to Resistive-wall Wake, PRST-AB, 7, 034402 ( 2004 ).}
\bibitem{5} {D.Whittum, Beam Breakup with Tune Chirp for an Arbitrary Wake Field, SLAC-PUB-7816 ( 1998 ).}
\bibitem{6} {P.B.Wilson, Introduction to Wakefields and Wake Potentials, SLAC-PUB-4547 ( 1989 ).}
\bibitem{8} {A.Chao, The Physics of Collective Beam Instability in High Energy Accelerators, Wiley, New York ( 1993 ).}
\end{thebibliography}
\end{document}